\documentclass[prb,twocolumn,aps,superscriptaddress,showpacs,floatfix]{revtex4}

\usepackage{amsmath}
\usepackage{amssymb}
\usepackage{amsbsy}
\usepackage{graphicx}
\usepackage{color}
\usepackage{epstopdf}
\usepackage{mathrsfs}
\usepackage{hyperref}
\usepackage{braket}
\usepackage{graphics}
\usepackage{stackengine}
\usepackage{verbatim}

\begin{document}
	
	\title{Non-Hermitian Aubry-Andr\'{e} model with Power-Law Hopping}
	
	\author{Zhihao Xu}
	\affiliation{Institute of Theoretical Physics and State Key Laboratory of Quantum Optics and Quantum Optics Devices, Shanxi University, Taiyuan 030006, China}
	\affiliation{Beijing National Laboratory for Condensed Matter Physics, Institute of Physics, Chinese Academy of Sciences, Beijing 100190, China}
	\affiliation{Collaborative Innovation Center of Extreme Optics, Shanxi University, Taiyuan 030006, China}
	
	\author{Xu Xia}
	\email{xiaxu14@mails.ucas.ac.cn}
	\affiliation{Chern Institute of Mathematics and LPMC, Nankai University, Tianjin 300071, China}
	
	\author{Shu Chen}
	\email{schen@iphy.ac.cn}
	\affiliation{Beijing National Laboratory for Condensed Matter Physics, Institute of Physics, Chinese Academy of Sciences, Beijing 100190, China}
	\affiliation{School of Physical Sciences, University of Chinese Academy of Sciences, Beijing, 100049, China}
	\affiliation{Yangtze River Delta Physics Research Center, Liyang, Jiangsu 213300, China}

	\begin{abstract}
		
		We study a non-Hermitian AA model with long-range hopping, $1/r^a$, and different choices of quasiperiodic parameters $\beta$ to be a member of the metallic mean family. We find that when the power-law exponent is in the $a<1$ regime, the system displays a delocalized-to-multifractal (DM) edge in its eigenstate spectrum. For the $a>1$ case, a delocalized-to-localized (DL) edge exists, also called the mobility edge. While a striking feature of the Hermitian AA model with long-range hopping is that the fraction of delocalized states can be obtained from a general sequence manifesting a mathematical feature of the metallic mean family, we find that the DM or DL edge for the non-Hermitian cases is independent of the mathematical feature of the metallic mean family. To understand this difference, we consider a specific case of the non-Hermitian long-range AA model with $a=2$, for which we can apply the Sarnak method to analytically derive its localization transition points and the exact expression of the DL edge. Our analytical result clearly demonstrates that the mobility edge is independent of the quasi-periodic parameter $\beta$, which confirms our numerical result. Finally, an optical setup is proposed to realize the non-Hermitian long-range AA model.
		
	\end{abstract}
	
	\maketitle
	\section{Introduction} \label{sec1}
	Quasicrystals exhibit an intermediate localization feature between fully periodic systems and fully disordered media. For the Anderson model with fully disordered media, an infinitesimal random potential results in localization in both one- and two-dimensional (1D and 2D) systems, whereas the mobility edges exist in 3D cases, which separate extended and localized single-particle states in the energy spectra \cite{Anderson,Anderson79}. The situation is changed in quasi-periodic cases. A paradigmatic example of a 1D quasicrystal system is the Aubry-Andr\'{e} (AA) model with nearest-neighbor hopping, which has been experimentally realized by the ultracold atomic technique in bichromatic optical lattices \cite{Roati,Deissler,Schreiber,Bordia}. A typical feature of the AA model is that above a finite critical quasi-periodic amplitude, all the eigenstates change from extended to localized, which is determined by its self-dual property \cite{Aubry,Suslov,Wilkinson}.
	
	Beyond nearest-neighbor hopping, breaking the self-duality of the AA model leads to the emergence of energy-dependent mobility edges, such as the system in shallow lattices \cite{Scherg,Hepeng,Diener,Ancilotto,Holthaus} or with exponential decay hopping \cite{Biddle}. Self-duality is also lost in the quasicrystals with power-law hoppings or interactions ($\propto 1/r^a$) \cite{Biddle1}, which have been particularly interesting since power-law interactions emerge in many systems, such as dipole-dipole interactions $(\propto 1/r^3)$ in polar molecules \cite{Gorceix,Moses}, Rydberg atoms \cite{Saffman}, nitrogen-vacancy centers \cite{Waldherr}, and nuclear spins in condensed-matter systems \cite{Kaiser}. The tunable power-law interactions can be realized in laser-driven ions ($0<a<3$) \cite{Richerme,Jurcevic}, which induce long-range exchange between the synthetic lattices resulting in power-law hoppings. The effect of power-law hoppings in the quasi-periodic AA model has been studied recently \cite{Shlyapnikov,NRoy}. One has shown the localization properties of the long-range AA model are characterized by the coexistence of localized (multifractal) states with delocalized states for $a>1$ ($a<1$) and by the emergence of a quasiperiodic parameter-dependent ladder of intermediate regimes in which the eigenstate blocks become localized or multifractal \cite{Shlyapnikov,NRoy}. By choosing a broader class of irrational Diophantine numbers, referred to as the metallic mean family, one shows the relation of the fraction of delocalized eigenstates and the irrational Diophantine numbers in the intermediate regimes \cite{NRoy}.
	
	On the other hand, non-Hermitian systems sparked a great interest both in experimental and theoretical fields \cite{Bender,Bender1,Hatano1,Hatano2,Hatano3,SLongi1,SLongi2,SLongi3,HuiJiang,HuitaoShen,Shunyu1,Shunyu2,Zongping,Lee1,Lee2,Harari,Parto,Schomerus18,Schomerus13,Schomerus15,Schomerus152,Schomerus2020,Schomerus2021,LiuZhouChen,HengyunZhou,Ruter,LiangFeng,Regensburger,Joglekar,XiangZhan,LeiXiao,Zeuner,YongXu,Okuma,LinhuLi,Kunst,Takata,MPan,Nakagawa,Hamazaki,Yamamoto,Ashida,Kawabata,Kawabata2,LihongZhou,Xuepeng1,Xuepeng2,Xuepeng3,XuePeng4,XuePeng5,XuePeng6,ZhihaoXu1,ZhihaoXu2,Linhu1,Longwen1}. Striking feathers are the failure of the bulk-boundary correspondence \cite{HuitaoShen,Shunyu1,Shunyu2,Zongping,Lee1,Takata}, the non-Hermitian skin effect \cite{SLongi1,HuiJiang,Shunyu1,Shunyu2,Zongping}, the sensitivity of the spectra on boundary conditions \cite{Hatano2,Kunst}, and the non-Hermitian-induced topology \cite{Takata,MPan}. Recently, the interplay of non-Hermiticity and disorder has brought a new perspective of the localization properties \cite{Hatano1,Hatano2,Hatano3,SLongi2,SLongi3,HuiJiang,LiuZhouChen,Zongping,ZhihaoXu4,ZhihaoXu3,SLonghi4,Longwen,YLiu2020,XuXia,YanxiaoLiu2021,YuCheng2021,TongLiu1,Longwen1}. The famous Hatano-Helson model describing the interplay of the random disorder and the nonreciprocal hopping in 1D lattices displays a finite metal-insulator transition \cite{Hatano1,Hatano2,Hatano3,Zongping}. One found that the coincidence of the metal-insulator phase transition point with the $\mathcal{PT}$ symmetry breaking point is of topological nature for the $\mathcal{PT}$ symmetrical extension of the AA models both in 1D \cite{SLongi2,SLongi3,HuiJiang,ZhihaoXu4,LiuZhouChen} and 2D systems \cite{ZhihaoXu3}. The non-Hermitian Maryland model shows a localization-delocalization phase transition via topological mobility edges in the complex energy plane \cite{SLonghi4,Longwen}, while there are not extended states in its Hermitian version \cite{Grempel,Simon}. These surprising results in non-Hermitian disorder fields inspire us to study the non-Hermitian effect in long-range AA models.
	
	In this paper, we study a non-Hermitian AA model with power-law hopping, and its quasi-periodic parameter is set to be a member of the metallic mean family, with special attention on the golden mean, silver mean, and bronze mean. Similar to the Hermitian cases \cite{Shlyapnikov,NRoy}, the non-Hermitian long-range AA model possesses the mixed regime where the multifractal (localized) states coexist with the delocalized ones for $a<1$ ($a>1$). However, our numerical results imply that the delocalized-to-multifractal (DM) and delocalized-to-localized (DL) edges are independent of the choice of irrational Diophantine numbers. We especially consider the non-Hermitian long-range AA model with $a=2$, for which we can apply the Sarnak method to obtain the exact expression of the DL edge. The analytical result for the DL edge clearly indicates that the mobility edge is independent of the quasi-periodic parameter $\beta$, which shows obviously different features from the Hermitian long-range AA. Such a non-Hermitian AA model with power-law hopping can be realized by an optical setup.
	
	This paper is organized as follows. In Sec. \ref{sec2}, we describe the Hamiltonian of the non-Hermitian AA model with power-law hopping and metallic means. In Sec. \ref{sec3}, we apply the fractal dimension to study the localization features of the eigenstates both for $a<1$ and $a>1$ regimes. We numerically and analytically display the distinct properties for non-Hermitian cases that the fraction of delocalized eigenstates is independent of the quasi-periodic parameter $\beta$. In Sec. \ref{sec4}, we propose an experimental scheme to realize such a non-Hermitian long-range model. Then we conclude in Sec. \ref{sec5}.
	
	\section{Model and Hamiltonian} \label{sec2}
	
	We consider a non-Hermitian AA model with power-law hopping, which can be described by the Hamiltonian
	\begin{equation}\label{eq1}
		\hat{H} = -J\sum_{j \ne i}\frac{1}{|j-i|^a}|j \rangle \langle i| + \sum_{j} \lambda_j |j\rangle \langle j|,
	\end{equation}
	with
	\begin{equation}\label{eq2}
		\lambda_j = \lambda e^{i2\pi\beta j},
	\end{equation}
	where $|j\rangle$ denotes the state in which the excitation is localized at the $j$th lattice site, and the hopping amplitude between sites $j$ and $j$ is $J/|j-i|^a$. We set $J=1$ as an energy unit. The complex on-site potential is characterized by its amplitude $\lambda$, and the incommensurabilty $\beta$ chosen to be a irrational Diophantine number. The potential is the complexification of $\lambda_j=2\lambda^{\prime}\cos(2\pi\beta+\varphi)$ taking $\varphi=ih$, and the limits $\lambda^{\prime}\to 0$, $h\to \infty$, keeping $\lambda^{\prime}e^{h}=\lambda$ finite \cite{SLongi2,SLongi3}. Notice that due to $\lambda_{-j}=\lambda_j^*$, the non-Hermitian model displays $\mathcal{PT}$ symmetry. Supposing that the eigenstate of a single particle in our system is given by $|\psi\rangle=\sum_{j}\psi_j |j\rangle$, we can obtain the eigenvalue equation as follows:
	\begin{equation}\label{eq3}
		-\sum_{n=1}^{\infty} \frac{1}{n^a}(\psi_{j-n}+\psi_{j+n})+\lambda_j\psi_j=E\psi_j,
	\end{equation}
	where $\psi_j$ is the amplitude of the particle wave function at the $j$-th site, $n=|j-i|$, and $E$ is the single-particle eigenvalue.
	
	The solution of Eq. (\ref{eq3}) is closely related to $a$ and the structure of the potential $\lambda_j$ for both non-Hermitian and Hermitian cases. For non-Hermitian cases, in the $a\gg 1$ limit \cite{SLongi3}, it has been verified that the metal-insulator transition emerges at $\lambda=J$, which also corresponds to $\mathcal{PT}$ symmetry breaking. When $\lambda<J$, all the eigenstates are delocalized, and its energy spectra are real numbers, while for $\lambda>J$, all the eigenstates are localized, and its energies become complex. For the Hermitian case, in addition to the fully localized and delocalized phases, one displays a coexistence of localized (multifractal) states with delocalized states for $a>1$ ($a<1$), where the fraction of delocalized states can be obtained from a general sequence manifesting the mathematical properties of the metallic mean family of the irrational Diophantine numbers \cite{Shlyapnikov,NRoy}. 
	
	To obtain the metallic mean family of the irrational Diophantine number, it is useful to consider a generalized $\kappa$-Fibonacci sequence,
	\begin{equation}\label{eq4}
		F_{\nu+1}=\kappa F_{\nu}+F_{\nu-1},
	\end{equation}
	with $F_0=0$ and $F_1=1$. The irrational number $\beta$ can be obtained by the limit $\beta=\lim_{\nu\to\infty} F_{\nu-1}/F_{\nu}$ with $\kappa=1,2,3,\cdots$, yielding the metallic mean family, such as the golden mean ($\beta_g=(\sqrt{5}-1)/2$) for $\kappa=1$, the silver mean ($\beta_s=\sqrt{2}-1$) for $\kappa=2$, the bronze mean ($\beta_b=(\sqrt{13}-3)/2$) for $\kappa=3$, and so on. In this paper, we take the system size $L=F_{\nu}$ and the rational approximation $\beta=F_{\nu-1}/F_{\nu}$ under periodic boundary conditions when numerically diagonalizing the non-Hermitian long-range quasicrystal model defined in Eq. (\ref{eq1}).
	
	\section{Localization transition in non-Hermtiain long-range AA model} \label{sec3}
	
	\begin{figure}[tbp]
		\begin{center}
			\includegraphics[width=.5 \textwidth] {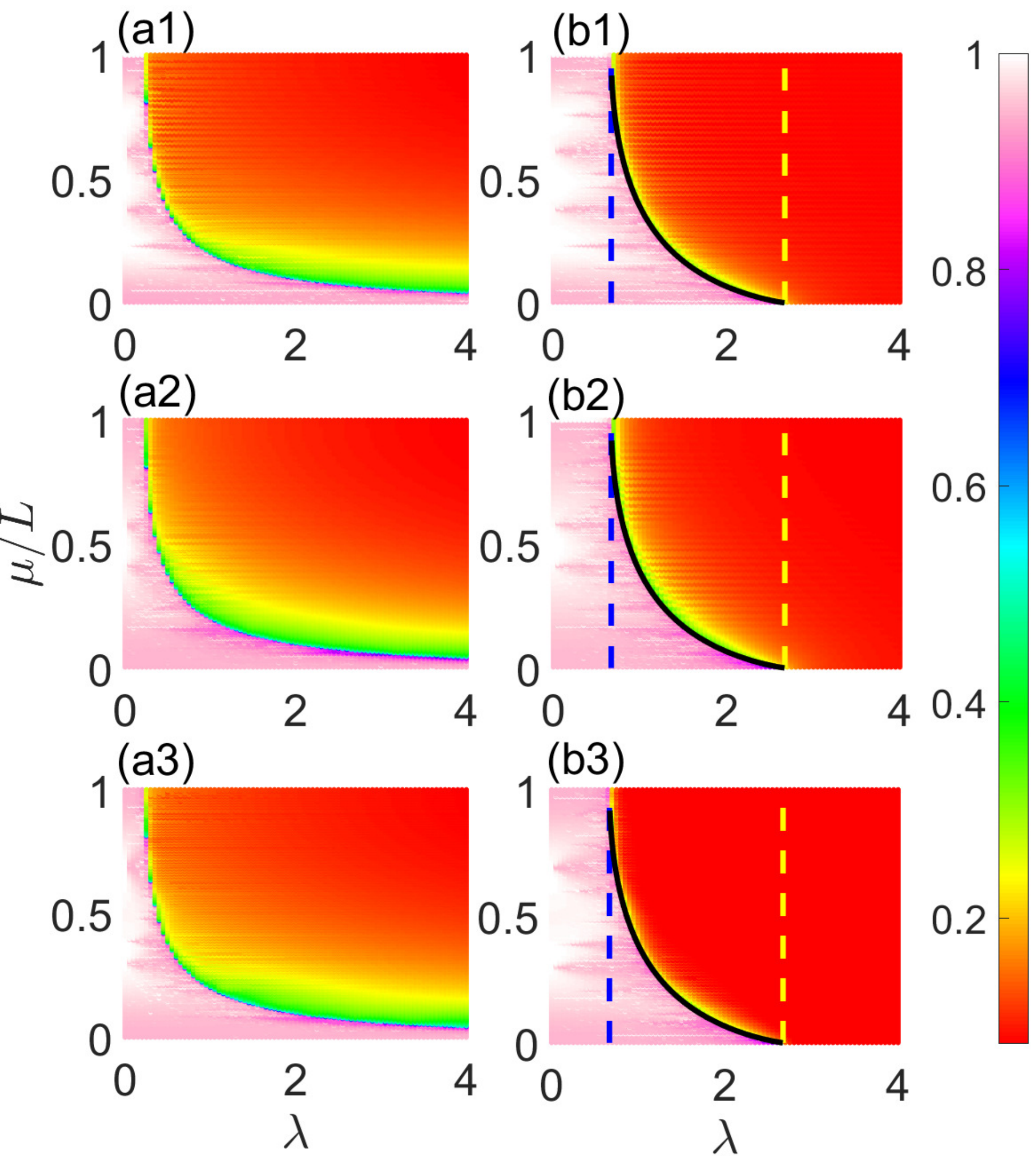}
		\end{center}
		\caption{(Color online) Fractal dimension $D_2$ of different eigenstates as a function of the index $\mu$ of the corresponding real part of the eigenvalues and the quasi-periodic potential amplitude $\lambda$ with (a1)--(a3) $a=0.5$ and (b1)--(b3) $a=2.0$, respectively. The real part of the energies is ordered in ascending order. The top row corresponds to $L=2584$, $\beta_g=1597/2584$, and $l=4$, the center row is for $L=2378$, $\beta_s=985/2378$ and $l=2$, and the bottom row is for $L=3927$, $\beta_b=1189/3927$, $l=3$, respectively. In (b1)--(b3), the blue and yellow dashed lines, respectively, represent $\lambda_{c1}\approx 0.6679$ and $\lambda_{c2} \approx 2.6714$, and the black solid lines denote the DL edges $E_c$ given by Eq. (\ref{eq12}).}\label{Fig1}
	\end{figure}

    \begin{figure}[tbp]
		\begin{center}
			\includegraphics[width=.5 \textwidth] {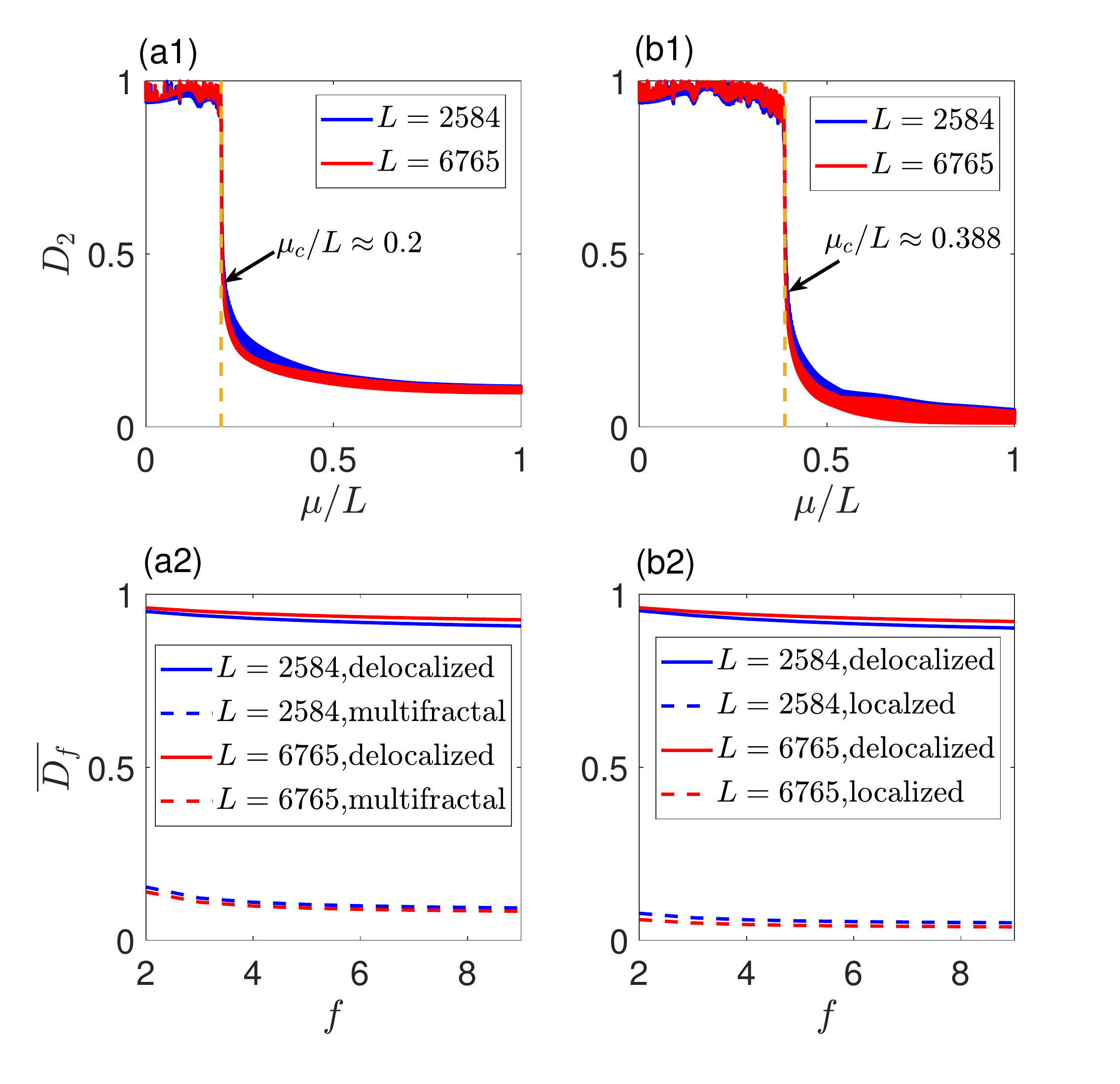}
		\end{center}
		\caption{(Color online) (a1)--(b1) $D_2$ versus the index $\mu/L$ of the real part of the eigenenergies for $\lambda=1.0$. The real part of the energies is ordered in ascending order. The dashed lines represent the energy indexes of the localization transitions $\mu_c/L$. (a2)--(b2) $\overline{D_f}$ as a function of $f$ for different $L$ and $\lambda=1.0$. $\overline{D_f}$ is calculated by averaging over the fraction of delocalized and the fraction of multifractal/localized eigenstates. The left column is for $a=0.5$ and the right one is for $a=2.0$, respectively.  Here, for $L=2584$, we choose $\beta_g=1597/2584$ and $l=4$, and for $L=6765$, we choose $\beta_g=4181/6765$, $l=5$, respectively.}\label{Fig2}
	\end{figure}

	One way of discerning between extended, localized, and multifractal states is given by the analysis of the eigenstates' fractal dimensions $D_f$ \cite{Chhabra,Janssen,Huckestein,Cuevas}. For an extended (localized) state, $D_f=0$ ($D_f=1$), whereas for a multifractal state, $D_f \in (0,1)$ and exhibits a dependence of $f$. To determine the fractal dimension, one divides the system of $L$ sites into $L_s=L/l$ boxes, and each box has $l$ sites. The fractal dimension is defined as
	\begin{equation}\label{eq5}
		D_f = \lim_{L_s\to \infty} \frac{1}{1-f} \frac{\ln{\left[ \sum_{p=1}^{L_s}(\mathcal{I}_p)^f \right]}}{\ln{L_s}},
	\end{equation} 	
	where $\mathcal{I}_p=\sum_{j\in p}|\psi^{(\mu)}_j|^2$ corresponding to the probability of detecting inside the $p$th box for the $\mu$th normalized eigenstate $|\psi^{(\mu)}\rangle$ with $\sum_{j=1}^{L} |\psi_j^{(\mu)}|=1$, and the corresponding eigenvalue $E_{\mu}$. Figures \ref{Fig1}(a1) and \ref{Fig1}(b1) show the fractal dimensions $D_2$ of the non-Hermitian long-range AA model as a function of $\lambda$ for all the eigenstates with the real part of eigenvalues in ascending order, when the non-Hermitian parameter is fixed at $\beta_g=1597/2584$ for $a=0.5$ and $2.0$, respectively. As seen from Fig. \ref{Fig1}(a1) with $a=0.5$, it displays a DM edge in the energy spectrum, where the fraction of extended states decreases and the multifractal states emerge with the increase of $\lambda$. Figure \ref{Fig2}(a1) shows the fractal dimension $D_2$ for different eigenstates $|\psi^{(\mu)}\rangle$ with different $L$, $\lambda=1.0$, and $a=0.5$, and the real part of the corresponding eigenvalues $E_{\mu}$ being ordered in ascending order. The DM edge emerges at $\mu_c/L\approx 520/2584 \approx 0.2$ for $\lambda=1.0$. When $\mu_c<520$, the corresponding eigenstates are extended with $D_2 \approx 1$. For $\mu_c>520$ for $\lambda=1.0$, the corresponding eigenstates show the multifractal feature with $D_2$ being finite values. However, for $a>1$, one can observe a DL edge, also called a mobility edge, in its energy spectrum. Figure \ref{Fig1}(b1) with $a=2.0$ shows the appearance of the block of localized states with $D_2\to 0$ with the increase of $\lambda$. Also shown in Fig. \ref{Fig2}(b1), the fractal dimension $D_2$ with different $L$, $\lambda=1.0$, and $a=2.0$ displays a jump from $D_2 \to 1$ to $D_2 \to 0$ at $\mu_c/L \approx 0.388$. As shown in Figs. \ref{Fig2}(a1) and \ref{Fig2}(b1), we can see the fractal dimensions $D_2$ are independent on the system size $L$.
	
	To further clarify the existence of mulitifractality in the regime $a<1$, we numerically calculate the average fractal dimension $\overline{D_f}$ for the regions with different localization features as a function of $f$ with $\lambda=1.0$ and the modulation parameter being $\beta_g$ for different $L$ shown in Figs. \ref{Fig2}(a2) ($a=0.5$) and \ref{Fig2}(b2) ($a=2.0$). Here, $\overline{D_f}$ represents $D_f$ averaged over the fraction of delocalized and the fraction of nondelocalized eigenstates. In Figs. \ref{Fig2}(a2) and \ref{Fig2}(b2), $\overline{D_f}$ averaged over the fraction of eigenstates with the indexes being less than $\mu_c$ are close to $1$ for different $f$, which implies these states are delocalized. $\overline{D_f}$ averaged over the fraction of eigenstates with the indexes being larger than $\mu_c$ are finite values and show a nontrivial dependence on $f$ for $a=0.5$, whereas $\overline{D_f}$ approach to $0$ and display almost no dependence on $f$ for $a=2.0$. It indicates that these states are multifractal for $a=0.5$ and localized for $a=2.0$. Also, we find $\overline{D_f}$ for different system sizes exhibit similar behaviors. Our numerical results imply that similar to the Hermitian cases, the non-Hermitian cases show a DM edge for $a<1$ and a DL edge for $a>1$.
	
	We also apply the Simon-Spencer theorem \cite{BSimon,book} to analytically discuss the absence of localized states for $a<1$ (for some details, see Appendix A). By performing the Fourier transformation,
	\begin{equation}\label{eq6}
		f(\theta)=\frac{1}{\sqrt{L}} \sum_j \psi_j e^{i 2\pi j\theta},
	\end{equation}
	we can obtain the dual equation of Eq. (\ref{eq3}),
	\begin{equation}\label{eq7}
		\lambda f(\theta+\tilde{\omega})=\left(E+\sum_m \frac{2}{m^a}\cos{m\theta}\right)f(\theta),
	\end{equation}
	with $\tilde{\omega}=2\pi\beta$, $\theta=2\pi\beta \tilde{j}$, $\tilde{j}$ being the site index in the dual space, and the potential of the system $\tilde{\lambda}_{\tilde{j}}=-\sum_m 2/m^a\cos{(2\pi \beta m \tilde{j})}$ in its dual space. The dual-Hamiltonian matrix in the thermodynamic limit can be written as:
	\begin{equation}\label{eq8}
		\tilde{H}=\begin{bmatrix} \ddots & \ddots & & & & & & \\
			 & \tilde{\lambda}_{-\tilde{j}_1} & \lambda & & & & & \\
			 & & \tilde{\lambda}_{-\tilde{j}_1+1} & \lambda & & & & \\
			 & & & \ddots & \ddots & & & \\
			 & & & & \tilde{\lambda}_{\tilde{j}_2-1} & \lambda & & \\
			 & & & & & \tilde{\lambda}_{\tilde{j}_2} & \lambda & \\
			 & & & & & & \ddots & \ddots & \\
			 \end{bmatrix}.
	\end{equation}
	The Simon-Spencer theorem says that if there are two series of monotonically increasing positive numbers, $\{\tilde{k}_s\}_{s=1}^{\infty}$ and $\{\tilde{k}^{\prime}_s\}_{s=1}^{\infty}$ to make $\sum_{s=1}^{\infty}\frac{1}{|\tilde{\lambda}_{-\tilde{k}_s}|}<\infty$ and $\sum_{s=1}^{\infty}\frac{1}{|\tilde{\lambda}_{\tilde{k}^{\prime}_s}|}<\infty$. Then the new block-diagonal dual-Hamiltonian
	\begin{equation}\label{eq9}
		\tilde{H}^{\prime}=\begin{bmatrix} \ddots & \ddots & & & & & & \\
			& \tilde{\lambda}_{-\tilde{k}_s} & 0 & & & & & \\
			& & \tilde{\lambda}_{-\tilde{k}_s+1} & \lambda & & & & \\
			& & & \ddots & \ddots & & & \\
			& & & & \tilde{\lambda}_{\tilde{k}^{\prime}_s-1} & 0 & & \\
			& & & & & \tilde{\lambda}_{\tilde{k}^{\prime}_s} & \lambda & \\
			& & & & & & \ddots & \ddots & \\
		\end{bmatrix}
	\end{equation}
	and the original dual-Hamiltonian $\tilde{H}$ possess the same absolutely continuous spectrum, whereas the existence of $\{\tilde{k}_s\}_{s=1}^{\infty}$ and $\{\tilde{k}^{\prime}_s\}_{s=1}^{\infty}$ ensures the absence of the absolutely continuous spectrum for $\tilde{H}^{\prime}$. For $0<a<1$, the dual potential $\tilde{\lambda}_{\tilde{j}}$ is unbounded, and one can always find out the two series $\{\tilde{k}_s\}_{s=1}^{\infty}$ and $\{\tilde{k}^{\prime}_s\}_{s=1}^{\infty}$ to keep $\sum_{s=1}^{\infty}\frac{1}{|\tilde{\lambda}_{-\tilde{k}_s}|}<\infty$ and $\sum_{s=1}^{\infty}\frac{1}{|\tilde{\lambda}_{\tilde{k}^{\prime}_s}|}<\infty$. Thus, our dual-Hamiltonian $\tilde{H}$ does not have the absolutely continuous spectrum, which means there are no localized states in the real space for our model in the $0<a<1$ regime. This proof is also suited for the Hermitian cases.
	
	\begin{figure}[tbp]
		\begin{center}
			\includegraphics[width=.5 \textwidth] {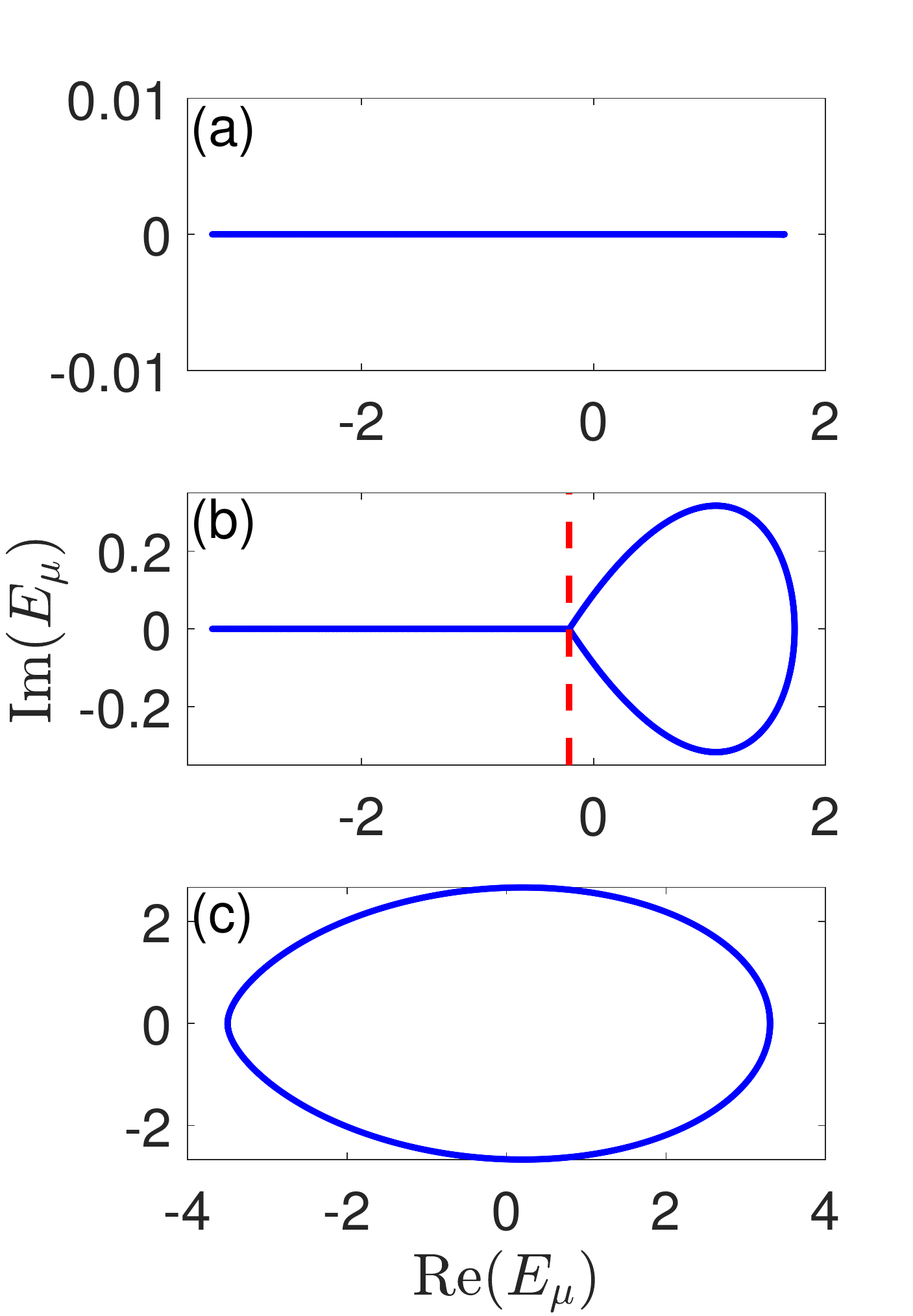}
		\end{center}
		\caption{(Color online) Energy spectrum $E_{\mu}$ of $\hat{H}$ for $\beta_g=1597/2584$ and $a=2$ with (a) $\lambda=0.5$, (b) $\lambda=1.0$, and (c) $\lambda=3.0$, respectively. The dashed line in (b) denotes the mobility edge $E_c\approx -0.2091$ corresponding to $\mu_c=1002$ for $\lambda=1.0$ and $L=2584$.}\label{Fig3}
	\end{figure}

	In the Hermitian cases, the fraction of the delocalized part depends on the choice of quasi-periodic parameter $\beta$. However, as shown in Fig. \ref{Fig1} for $\beta$ being chosen as the different metallic mean family of irrational numbers, one can compare the numerical results with different quasi-periodic parameters $\beta$ for the same $a$, and easily find that, unlike the Hermitian case, the fraction of delocalized eigenstates is independent of $\beta$ in the non-Hermitian cases. To analytically verify the property, as a concrete example, we study the localization features of the non-Hermitian long-range AA model with $a=2$. The dual equation for $a=2$ is given as follows:
	\begin{equation}\label{eq10}
		\lambda f(\theta+\tilde{\omega})=(E+\frac{\theta^2}{2}-\pi \theta+\frac{\pi^2}{3})f(\theta).
	\end{equation}
	According to the Sarnak method \cite{Sarnak,ZhihaoXu3} (for some details on the Sarnak method, see Appendix B), we define a characteristic function as follows:
	\begin{align}\label{eq11}
		G(E) &=\frac{1}{2\pi}\int_{0}^{2\pi} \ln{\left|E+\frac{\theta^2}{2}-\pi \theta+\frac{\pi^2}{3}\right|} \notag \\
		&= -2-\ln{2}+\frac{u_{E1}\ln{u_{E1}}+u_{E2}\ln{u_{E2}}}{\pi},
	\end{align}
	with $u_{E1}=\pi+\sqrt{\pi^2/3-2E}$ and $u_{E2}=\pi-\sqrt{\pi^2/3-2E}$. The Sanark method says (i) when $G(E)=\ln{|\lambda|}$, the spectrum is dense with the localized eigenstates, and the corresponding eigenvalue $E$ is a complex value and (ii) when $\{G(E)>\ln{|\lambda|}\}\cap U_{E}$, it corresponds to a dense set of $E$ with the extended eigenstates, where $U_{E}=[-\pi^2/3,\pi^2/6]$ is the set of the spectrum such that $E+\theta^2/2-\pi\theta+\pi^2/3=0$ for some $\theta$, and $E$ is a real value. For the eigenenergies in the region $U_E$, we find $G(E)\in [2\ln{\pi}-\ln{2}-2,2\ln{\pi}+\ln{2}-2]$. Hence, two delocalization-localization transition points exist at $\lambda_{c1}=\exp{(2\ln{\pi}-\ln{2}-2)} \approx 0.6679$ and $\lambda_{c2}=\exp{(2\ln{\pi}+\ln{2}-2)} \approx 2.6714$, which are labeled by blue and yellow dashed lines, respectively, shown in Figs. \ref{Fig1}(b1)--\ref{Fig1}(b3). When $\lambda<\lambda_{c1}$, all the eigenstates are extended, and the corresponding energy spectrum is entirely real, which is shown in Fig. \ref{Fig3}(a) with $\lambda=0.5$ and $\beta_g=1597/2584$. For $\lambda>\lambda_{c2}$, all the eigenstates are localized, and the corresponding energy spectrum becomes totally complex shown in Fig. \ref{Fig3}(c) with $\lambda=3.0$ and $\beta_g=1597/2584$. The mobility edge emerges in the region $\lambda \in [\lambda_{c1},\lambda_{c2}]$. In this intermediate regime, the DL edge $E_c$ is a real number that satisfies
	\begin{equation}\label{eq12}
		G(E_c)=\ln{\lambda},
	\end{equation}
	which is marked by the black solid lines in Figs. \ref{Fig1}(b1)--\ref{Fig1}(b3). As shown in Fig. \ref{Fig3}(b) with $\lambda=1.0$ and $\beta_g=1597/2584$, when $\mathrm{Re}(E_{\mu})<E_c\approx -0.2091$, the eigenenergies are real values, while when $\mathrm{Re}(E_{\mu})>E_c$, the eigenenergies become complex values. Here, $E_c\approx -0.2091$ for $\lambda=1$ corresponds to $\mu_c=1002$, which is marked by the red dashed line in Fig. \ref{Fig3}(b). According to the numerical and analytical results, we find $\lambda_{c1}$ is also the $\mathcal{PT}$ symmetry broken point corresponding to the point emerging in the complex eigenenergies, and the DL edges are independent on $\beta$. In Appendix C, we show the relation of $\mathcal{PT}$-symmetry transition and the DM transition. Our results imply that when the quasi-periodic parameter $\beta$ is chosen as the different metallic mean family of irrational Diophantine numbers, the fraction of delocalized states shall display the same feature.	
	
	\section{Optical realization of non-Hermitian long-range AA model} \label{sec4}
	
	\begin{figure}[tbp]
		\begin{center}
			\includegraphics[width=.5 \textwidth] {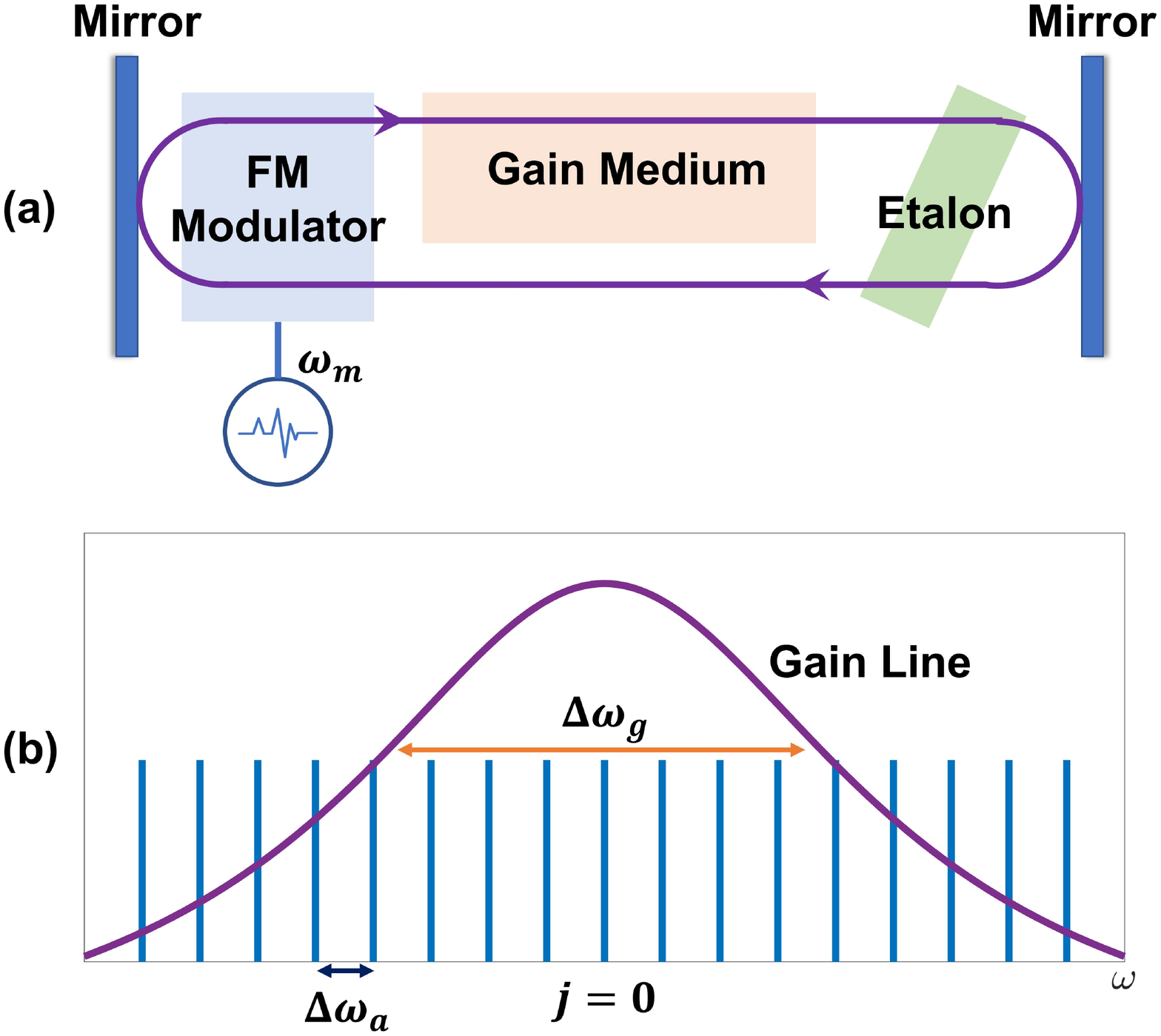}
		\end{center}
		\caption{(Color online) (a) Schematic of the FM multi-mode laser. (b) The axial spectral modes of the laser cavity spaced by $\Delta \omega_a$ realized a 1D non-Hermitian lattice with the long-range hopping. The full width at half maximum of the gain line is $\Delta \omega_g$ with $\Delta \omega_g \gg \Delta \omega_a$. }\label{Fig4}
	\end{figure}

	In the experiment, one can implement the non-Hermitian long-range AA model in an optical setup by applying the frequency-modulated (FM) multi-mode laser \cite{Kuizenga1,Kuizenga2} shown in Fig. \ref{Fig4}(a). The optical setup is done with a standard Fabry-Perot laser cavity with the axial modes spaced by $\Delta \omega_a$ undergoing the FM modulator, the gain medium, and the etalon. The light circulates back and forth between the two end mirrors of the cavity, and the round-trip number in the cavity is marked by $t$, which can be normalized to physical time. When the light finishes a round trip, the spectral components $\psi_j$ of cavity axial modes can be given \cite{Haus}
	\begin{align}\label{eq13}
		\psi_j(t+1)&=\psi_j(t)+\{ \delta\psi_j(t)\}_{\mathrm{loss}}+\{ \delta \psi_j(t)\}_{\mathrm{modul}} \notag \\
		&+\{\delta \psi_j(t)\}_{\mathrm{gain}}+\{ \delta \psi_j(t)\}_{\mathrm{etalon}}.
	\end{align}
	Here, $\{ \delta\psi_j(t)\}_{\mathrm{loss}}=-\gamma_c \psi_j(t)$ represents the losses of the cavity with loss rate $\gamma_c$ per round-trip. $\{ \delta \psi_j(t)\}_{\mathrm{modul}}$ is the change from the FM modulator, which introduces a time-dependent phase shift $\Delta \varphi(t) = \Delta_{\mathrm{FM}} \sum_{n=1} \cos{(n\omega_m t)}/n^a$ to the incident light field with the amplitude $\Delta_{\mathrm{FM}}\ll 1$ and the frequency $\omega_m=\Delta \omega_a$. One has
	\begin{equation}\label{eq14}
		\{ \delta \psi_j(t)\}_{\mathrm{modul}}=i\frac{\Delta_{\mathrm{FM}}}{2}\sum_{n=1}\frac{1}{n^a}\left[\psi_{j}(t)e^{in\omega_m t}+\psi_{j}(t)e^{-in\omega_m t}\right].
	\end{equation}
	When the round-trip passes through the gain medium, one has\cite{Haus}
	\begin{equation}\label{eq15}
		\{\delta \psi_j(t)\}_{\mathrm{gain}}\approx \frac{g}{1+4j^2\omega_m^2/\Delta \omega_g^2} \psi_j(t),
	\end{equation}
	which is the small change of $\psi_j$ for a homogeneously broadened active material with a slow relaxation dynamics and wide gain bandwidth $\Delta \omega_g \gg \Delta \omega_a$ \cite{Kuizenga1} shown in Fig. \ref{Fig4}(b). Here, $g$ is the saturated gain. The change arising from etalon transmission is represented by $\{ \delta \psi_j(t)\}_{\mathrm{etalon}}$. We assume the etalon with the refractive index $\bar{n}$ and the thickness $\bar{L}$ is placed inside the cavity and slightly tilted from the normal incidence. For near-normal incidence, the spectral transmission of the etalon $T(\omega)=(1-R)/[1-R\exp{(i2\bar{n} \bar{L} \omega /c+i\phi)}]$ with the reflectance $R\approx (\frac{\bar{n}-1}{\bar{n}+1})^2$ of etalon facets, the speed of light in vacuum $c$, and the additional phase shift $\phi$ adjusted by slight etalon tilting. For a glass-air interface, $R\ll 1$,  we have $T(\omega)\approx 1-R+R\exp{(i2 \bar{n} \bar{L} \omega /c+i\phi)}$, and
	\begin{align}\label{eq16}
		\{ \delta \psi_j(t)\}_{\mathrm{etalon}} &=[T(j\omega_m)-1]\psi_j(t) \notag \\
		&=\left[-R+R e^{i(2\pi\beta j+\phi)}\right]\psi_j(t),
	\end{align}
	where $\beta=\omega_m/\Delta \omega_{et}$ and $\Delta \omega_{et}=\pi c/(\bar{n}\bar{L})$. After setting $\psi_{j}(t+1)-\psi_j(t)\approx (\partial \psi_j/\partial t)$ and $\psi_j(t)=\psi_j \exp{(-i\omega_m t)}$, one can obtain from Eq. (\ref{eq13}) \cite{Haus,Gordon}
	\begin{equation}\label{eq17}
		i\frac{\partial \psi_j}{\partial t}=-J\sum_{n=1}\frac{1}{n^a}(\psi_{j-n}+\psi_{j+n})+\lambda e^{i2\pi\beta j}\psi_j+i\mathcal{L}_j\psi_j,
	\end{equation}
	where $J=\Delta_{FM}/2$, $\lambda=R$, $\mathcal{L}_j=-\gamma_c-R+g/(1+4 j^2\omega_m^2/\Delta \omega_g^2)$, and $\phi=\pi/2$. In the broad gain-line limit, $\omega_m/\Delta \omega_g\to 0$, and setting $\gamma_c+R \approx g$, the spectral mode dynamics can emulate the non-Hermitian long-range AA model.
	

	\section{Conclusions} \label{sec5}
	The study of long-range quasicrystals in Hermitian systems exhibits rich localization phenomena. Recent fruitful achievements on the interplay of non-Hermiticity and quasicrystals inspire us to consider the non-Hermitian AA model with power-law hopping and the quasiperiodic parameter chosen to be a set of the metallic mean family. In this paper, we find in addition to the fully delocalized and localized regime, a coexistence of multifractal (localized) eigenstates with delocalized eigenstates for $a<1$ ($a>1$), which is similar to the Hermitian cases. Unlike the Hermitian long-range AA model, we analytically and numerically verify that the fraction of delocalized eigenstates is independent of the quasi-periodic parameter in the intermediate regime. We propose an optical setup by using the FM laser to realize the non-Hermitian long-range AA model. We believe that our study will motivate further studies on the exploration of localization properties in non-Hermitian long-range quasicrystals.
	
	\section*{Appendix A: Some details on Simon-Spencer theorem}
	
	The Simon-Spencer theorem is based on the trace class perturbation method \cite{book}, which is used to transform an infinite-dimensional matrix ($\hat{\mathcal{A}}$) to a direct sum of a series of finite-dimensional matrices ($\hat{\mathcal{B}}=\bigoplus_k \hat{b}_k$) by a series of perturbation operators. Under some kinds of trace class perturbation, one can guarantee such two matrices $\hat{\mathcal{A}}$ and $\hat{\mathcal{B}}$ have the same absolutely continuous spectrum if the trace class norm can be effectively controlled. 
	
	Applied to our case for $a<1$, a trace class perturbation corresponds to changing the left hopping amplitude from $\lambda$ to $0$ at some positions $\tilde{k}_s$ for the infinite-dimension matrix $\tilde{H}$ described by Eq. (\ref{eq8}) in the main text, and the trace class norm is then controlled by $\frac{1}{|\tilde{\lambda}_{\tilde{k}_s}|}$. By performing a series of trace class perturbations, the infinite system can be decomposed into an infinite number of small closed systems described by $\tilde{H}^{\prime}$ in the main text, naturally in which there are no extended states. Due to the unbounded dual potential $\tilde{\lambda}_{\tilde{j}}$, the trace class norms can be controlled by $\sum_{s=1}^{\infty} \frac{1}{|\tilde{\lambda}_{\tilde{k}_s}|} < \infty$ and $\sum_{s=1}^{\infty} \frac{1}{|\tilde{\lambda}_{-\tilde{k}^{\prime}_s}|} < \infty$ for two series of monotonically increasing positive numbers $\{\tilde{k}_s\}_{s=1}^{\infty}$ and $\{\tilde{k}^{\prime}_s\}_{s=1}^{\infty}$. According to the Simon-Spencer theorem, we can say that $\tilde{H}$ and $\tilde{H}^{\prime}$ have the same absolutely continuous spectrum. Hence, the dual equation $\tilde{H}$ does not have extended states.
	
	\section*{Appendix B: Some details on the Sanark method}
	The Sanark method can solve a class problems for quasicrystals with the eigenvalue equation described by:
	\begin{equation}\label{seq1}
		f(\theta+\tilde{\omega})=g(E,\theta)f(\theta),
	\end{equation}
where $E$ is a given value, $\theta=2\pi\beta j$, $\tilde{\omega}=2\pi\beta$, and $j$ is the site index. One can define the Lyapunov exponent as
	\begin{align}\label{seq2}
		\mathcal{L}(E) &= \lim_{m\to \infty}\frac{1}{m} \ln{\left|\frac{f(\theta+m\tilde{\omega})}{f(\theta)}\right|} \notag \\
		&= \lim_{m\to \infty} \frac{1}{m} \sum_{k=0}^{m-1} \ln{\left| g(E,\theta+k\tilde{\omega})\right|}.
	\end{align}
In the main text, $G(E)=\mathcal{L}(E) + \ln{\lambda}$. Then if the Lyapunov exponent $\mathcal{L}(E)=0$, the given value $E$ is the eigenvalue of Eq. (seq1) and the corresponding eigenstate is an extended state. If $\mathcal{L}(E) \ne 0$ and $g(E,\theta) \ne 0$ are always true, according to the ergodic theory that such $E$ is not the solution of the equation. For some $\theta$, $g(E,\theta)=0$, and when $\mathcal{L}(E)>0$, the given $E$ is the eigenvalue corresponding to a localized state.

	\section*{Appendix C: $\mathcal{PT}$ symmetry transition for $a<1$}
	\begin{figure}[tbp]
		\begin{center}
			\includegraphics[width=.5 \textwidth] {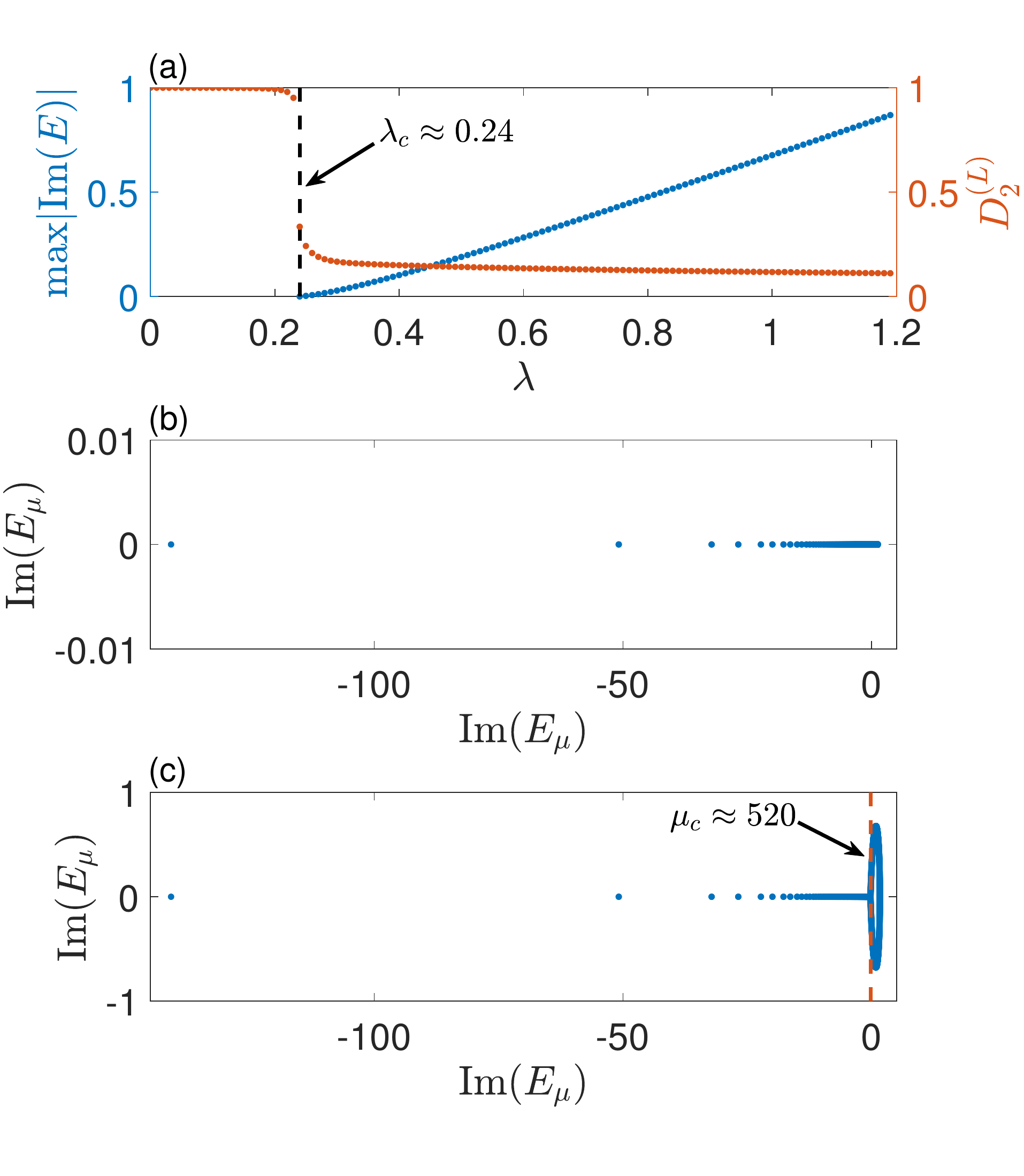}
		\end{center}
		\caption{(Color online) (a) The behavior of the maximum value of $|\mathrm{Im}(E)|$ and the fractal dimension of the $L$-th eigenstate $D_2^{(L)}$ as the functions of $\lambda$ for $a=0.5$ and $\beta_g=1597/2584$. The dashed line in (a) denotes the $\mathcal{PT}$ symmetry breaking point and the DM transition point. Energy spectrum $E_\mu$ of $\hat{H}$ for $a=0.5$ and $\beta_g=1597/2584$ with (b) $\lambda=0.1$ and (c) $\lambda=1.0$, respectively. The dashed line in (c) corresponds the DM edge with $\mu_c \approx 520$ for $\lambda=1.0$ and $L=2584$.}\label{Fig5}
	\end{figure}

	In this appendix, we consider the $\mathcal{PT}$ symmetry transition in the $a<1$ region. Figure \ref{Fig5}(a) shows the behavior of the maximum value of $|\mathrm{Im}(E)|$ and the fractal dimension of the $L$-th eigenstate $D_2^{(L)}$ as the functions of $\lambda$ for $a=0.5$ and $\beta_g=1597/2584$. Here, the real part of the eigenvalues is ordered in ascending order. According to the numerical calculation of fractal dimension for $a=0.5$ and $\beta_g=1597/2584$ shown in Fig. \ref{Fig1}(a1) in the main text, one can see that the DM transition point is denoted by the eigenstate with the maximum real part of the energy. As seen in Fig. \ref{Fig5}(a), the $\mathcal{PT}$ symmetry breaking point coincides with the DM phase transition point at $\lambda_c \approx 0.24$ for $a=0.5$. The energy spectrum for $a=0.5$ and $\lambda<\lambda_c$ is shown in Fig. \ref{Fig5}(b) ($\lambda=0.1$), where all the eigenvalues are real. When $\lambda>\lambda_c$, the complex energies emerge. For the case of $\lambda=1.0$ and $a=0.5$ [Fig. \ref{Fig5}(c)], the real-complex transition of the energy spectrum emerges at $\mu_c\approx 520$ which corresponds to the DM edge shown in the main text. In conclusion, similar to the case of $a>1$, we numerically verify the coincidence of the $\mathcal{PT}$ symmetry breaking point with the DM phase transition point for the long-range quasicrystals ($a<1$).

	\begin{acknowledgements}
		Z.X. thanks Linjie Zhang for helpful discussions. Z. X. is supported by the NSFC (Grants No. 11604188 and No. 12047571), Beijing National Laboratory for Condensed Matter Physics, and STIP of Higher Education Institutions in Shanxi under Grant No. 2019L0097. X.X. is supported by NanKai Zhide Foundation. S.C. is supported by NSFC under Grant No. 11974413 and the Strategic Priority Research Program of Chinese Academy of Sciences under Grant	No. XDB33000000. This work is also supported by NSF for Shanxi Province Grant No. 1331KSC.
		
	\end{acknowledgements}

\end{document}